# Infrared spectroscopy across scales in length and time at BESSY II

Alexander Veber[a,b], Ljiljana Puskar[b], Janina Kneipp[a], Ulrich Schade[b]

[a] Department of Chemistry, Humboldt-Universität zu Berlin, Brook-Taylor-Straße 2, 12489 Berlin, Germany

[b] Institute for Electronic Structure Dynamics, Helmholtz-Zentrum Berlin für Materialien und Energie GmbH, Albert-Einstein-Straße 15, 12489 Berlin, Germany

**Synopsis** Recent upgrade of the infrared beamline at BESSY II storage ring made improved characterization of molecules and materials at different length and time scales possible. The new nano-spectroscopy end-station based on a scattering type scanning optical microscope enables IR-imaging and spectroscopy with a spatial resolution better than 30 nm.

**Abstract** The infrared beamline at BESSY II storage ring was upgraded recently to extend capabilities of infrared microscopy. The end-stations available at the beamline are now facilitating improved characterization of molecules and materials at different length scales and time resolution. We report the current outline of the beamline and give an overview of the end-stations available. In particular, presented here are first results obtained with using a new microscope for nano-spectroscopy that was implemented. We demonstrate the capabilities of the scattering-type near-field optical microscope (s-SNOM) by investigation of cellulose microfibrils, representing nanoscopic objects of a hierarchical structure. It is shown that the s-SNOM coupled to the beamline allows to perform imaging with spatial resolution less than 30 nm and to collect infrared spectra from effective volume of less than $30 \times 30 \times 12$ nm$^3$. Potential steps for a further optimization of the beamline performance are discussed.

**Keywords:** Infrared beamline; BESSY II; nano-spectroscopy; cellulose microfibrils.

## 1. Introduction

The infrared (IR) beamline IRIS at the BESSY II storage ring was inaugurated in 2001 and it is currently the only infrared beamline available in Germany for national and international user groups(Peatman & Schade, 2001; Schade *et al.*, 2002). The design of the beamline allows to extract broadband high brilliance radiation spanning from 2 to 10000 cm$^{-1}$. In combination with the top-up operation mode of the BESSY II storage ring(Kuske *et al.*), this turns the beamline to stable radiation source that is well-suited well for a diverse range of different IR-spectroscopy applications.

The back-end and the end-stations are separated from the UHV front-end of the beamline, providing great flexibility in redistribution of the synchrotron radiation and modification of the end-stations. During more than 20 years of the operation the beamline was constantly developing and evolving to meet demands of the scientific community, hence offer unique, cutting-edge IR-spectroscopy experimental techniques to the users. Initially equipped with an Fourier-transform infrared (FTIR) spectroscopy and a micro-spectroscopy end-stations, the beamline was complemented by an

ellipsometer(Gensch *et al.*, 2003; Hinrichs *et al.*, 2003) and THz-imaging/spectroscopy set-ups (Schade *et al.*, 2004) in 2002-2004, instrumentation for microscopy observation of vibrational linear dichroism using polarization-modulated IR synchrotron radiation in 2006-2008(Schmidt *et al.*, 2006, 2008), followed by a dispersive single shot time-resolved spectrometer(Ritter *et al.*, 2019) in 2018-2020. During a recent upgrade in 2020-2023 the beamline was extended to fit an additional infrared scattering type scanning near-field optical microscope (s-SNOM). Moreover imaging capabilities for the micro-spectroscopy were extended. In this work, we present the current outline of the beamline and describe the available end-stations and their parameters. We discuss and demonstrate capabilities of the new nano-spectroscopy end-station in more detail. Finally, we discuss the ongoing equipment developments and further modernization of the beamline.

**2. Beamline design**

Figure 1 displays the current outline of the beamline. The front-end of the beamline has been described (Peatman & Schade, 2001; Schade *et al.*, 2002). In brief, a slotted mirror (M1) is used to extract the optical radiation originating for the homogeneous region of the bending magnet after the optical beam is refocused twice with two sets of cylindrical mirrors (M2-M3 and M4-M5). Thereafter the beam is collimated by a toroid mirror (M6) and can be redirected to the different end-stations. The beam is not divided in parts and entire optical radiation is delivered to the respective selected end-station.

The optical scheme extracts the infrared beam upwards to the top of the storage ring. Previously, all the end-stations were mounted on the monolithic concrete roof of the storage ring, which guaranteed high mechanical stability of the equipment. Due to the space constraints, the additional nano spectroscopy end-station would not have fitted on the tunnel roof, therefore the beamline was extended to the ground level of the storage ring experimental hall and an FTIR microscope was accommodated there (Figure 1, port 3). The vacuum and optical systems were modified to provide now four optical ports: three ports on the roof (Figure 1, ports 1,2, and 4) and one additional at the ground level (Figure 1, port 3). The location of the ports 1 and 2 remained unchanged, and they are currently used for spectroscopy and single-shot time-resolved spectroscopy end-stations, respectively. The micro-spectroscopy, previously also located on the roof, was relocated to the ground level (port 3), whereas the space at the roof was used for the nano-spectroscopy end-station (port 4). To re-collimate and deliver the IR beam to the port 3 at the ground level and the port 4 on the tunnel roof two additional pairs of toroidal mirrors and a couple of plane mirrors were added to the original optical scheme of the beamline. Each mirror reflects the light by 45°. The mirror pairs M7-M8 and M9-M10 re-collimate and resize the beam with a magnification factor of 0.6 and 1.6 for the ports 4 and 3, respectively. The magnification factors were chosen to fit the optical input of the corresponding nano- and micro-spectroscopy end-stations.

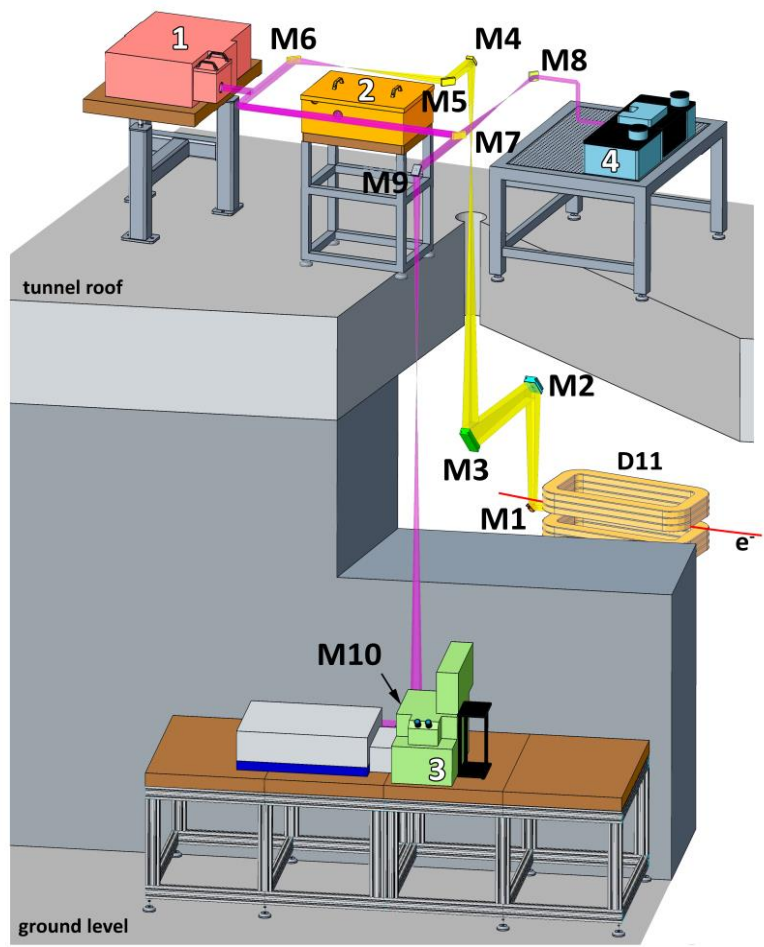

**Figure 1** Outline of the current IRIS beamline showing the state as of September 2023. The end-stations are marked as follows: 1. IR spectroscopy; 2. Single-shot time resolved IR spectroscopy; 3. IR micro-spectroscopy; 4. IR nano-spectroscopy.

## 3. End-stations of the beamline

### 3.1. IR macro-spectroscopy

Currently four end-stations are available to the users of the beamline. The information about the end-stations, the corresponding spectral ranges, available accessories and the measurement modes are listed in Table 1. The IR macro-spectroscopy end-station is based on Bruker Vertex 70v vacuum FTIR spectrometer (Bruker, Leipzig, Germany) and meant mostly for investigation of macroscopic samples. The measurements with this spectrometer can be performed in the spectral range 20-10000 cm$^{-1}$ in the incoherent multi-bunch synchrotron operational mode. A special low-alpha operational mode of the synchrotron results in coherent synchrotron radiation at sub-terahertz frequencies, which allows extend the spectral range at the IR beamline and to collect high-quality data in the spectral range down to 3 cm$^{-1}$ (Abo-Bakr *et al.*, 2003; Schade *et al.*, 2007; Puskar & Schade, 2016). This end-station provides the widest spectral range as well as the most variation of the sample environment, including wide temperature range from 4 to 800 K, controlled gas and humidity atmosphere, multiple measurement modalities suitable for the investigation of solid and liquid materials. In addition to steady state measurements the spectrometer also allows to perform time resolved experiment using the step scan and the rapid scan techniques. The end-station suits well for precise characterization of novel materials, the study of phase transitions, chemical reactions etc.

### 3.2. Single shot time resolved IR spectroscopy

A dedicated spectrometer was implemented for the investigation of non-reversible processes or processes with a slow recovery kinetics (Schade *et al.*, 2014; Ritter *et al.*, 2019), specifically in studies of protein structure and dynamics. The experimental setup exploits a home built dispersive IR spectrometer based on a Féry prism, which allows to collect single- shot spectra in the spectral range of 1000-1800 cm$^{-1}$ with the time resolution of about 5 µs. The current implementation of the end-station allows for the studies of photo induced conformational changes. The performance of the instrument was demostrated previously on both the irreversible activation of vertebrate rhodopsin and slow-cycling microbial actinorhodopsin systems initiated by a laser pulse of 532 nm (Ritter *et al.*, 2019).

**Table 1**  List of the end-stations available for user operation at the IRIS beamline and their main characteristics.

| End-station | Equipment used | Spectral range (cm$^{-1}$) | Spatial resolution @ λ=10 µm | Available modalities | Special sample environment |
|---|---|---|---|---|---|
| 1. IR macro-spectroscopy | Bruker Vertex 70V | 2-10000* | ≥ 1 mm | Transmission; Specular/ diffuse reflection; ATR; Polarimetry, including PEM; rapid and step-scan; Far-IR microscopy | Vacuum; T=4 - 800 K; Controlled gas/humidity atmosphere |
| 2. Single-shot spectroscopy | Dispersive Fery prism spectrometer | 1000-1800 | ~1 mm | 5 µs time resolution; transmission; Photoexcitation @ 532 nm. | |
| 3. IR micro-spectroscopy | Bruker Vertex 80 + Hyperion 3000 | 80-5000 | ~6 µm | Transmission; Reflection; ATR; Polarimetry, including PEM; FPA (64x64 pixels) hyperspectral imaging; rapid and step-scan | T= 77-500 K; Diamond compression cell |

| 4. IR nano-spectroscopy | Neascope scattering type near-field optical microscope | 600-2000** | ≥ 25 nm | Reflection; white light imaging; FTIR spectroscopy; PsHet imaging*** | Under development |

*the spectral range of 2-20 cm$^{-1}$ is available in low-alpha operational mode of the storage ring only

**detector limited and is planned to be extended to 330-4000 cm$^{-1}$ in early 2024

***PsHet is a laser source-based modality and available in the spectral range of 1675-1865 cm$^{-1}$

### 3.3. Upgrade of the micro-spectroscopy end-station

A Continuum microscope coupled to a Nexus 670 spectrometer (Nicolet, Madison, WI, USA), that had been in use over two decades, was replaced by a Hyperion 3000 IR microscope and a Vertex 80 FTIR spectrometer (Bruker Optics GmbH, Ettlingen, Germany). The new microscope ensures stable in time operation and the experiments in mid-infrared spectral range can be done with use of a single point mercury cadmium telluride (MCT) detector or an MCT 64x64 pixel focal plane array (FPA) detector, which allows to perform measurements faster over the larger areas of interest. The vibrational linear dichroism modality(Schmidt *et al.*, 2006) was successfully transferred to the new microscope. The step scan and the rapid scan techniques make the microscope suitable for time-resolved experiments.

The new microscope has an optical port for a Si-bolometer, which expands the spectral range available for the end station in the low wavenumber rage to about 80 cm$^{-1}$. The advantage of the synchrotron in experiments with diffraction-limited spatial resolution in the far-IR/THz spectral range is evident from Figure 2, comparing the signal intensity of the synchrotron and a Globar source measured through a 100x100 μm$^2$ aperture. The use of the synchrotron radiation results in about one order of magnitude benefit in comparison to the internal Globar source. An rms noise of better than 0.1% is achieved in the range from 150 to 550 cm$^{-1}$, as indicated by the 100% line, calculated as the ratio of two subsequently recorded spectra (Figure 2, bottom panel).

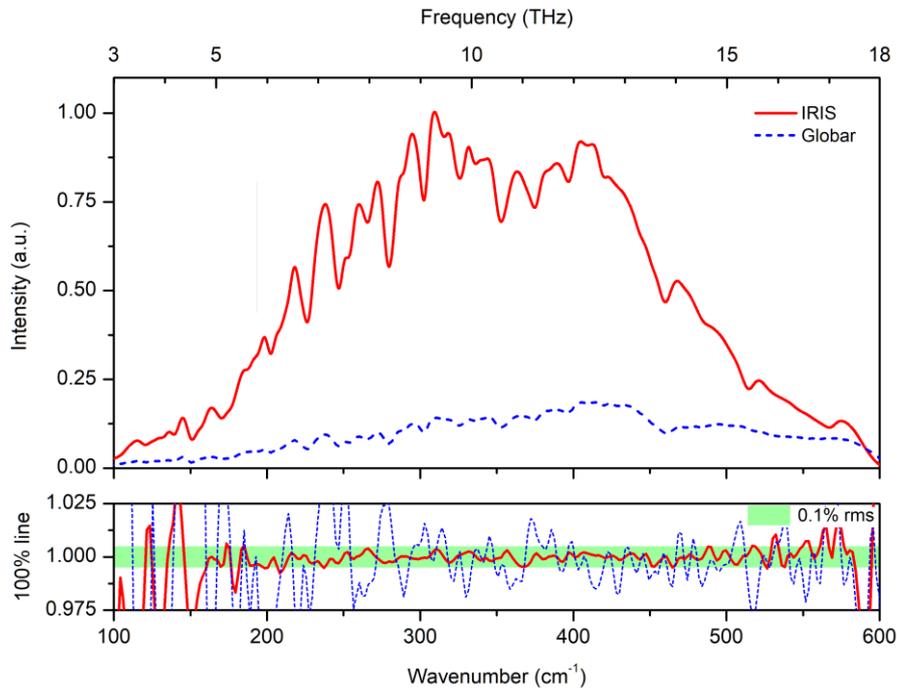

**Figure 2** Comparison of the fluxes in the far IR spectral range at the micro-spectroscopy end-station. Using the infrared synchrotron radiation the signal intensity is about one order of magnitude higher than using the internal Globar source. An rms noise of better than 0.1% is achieved from the synchrotron source in the range of 150-550 cm$^{-1}$, as indicated by the 100% line, the ratio of two subsequently recorded spectra. The signal is recorded with the square knife aperture set to 100x100 μm$^2$ using the Si-Bolometer.

### 3.4. Nano-spectroscopy end-station

The nano-spectroscopy end-station is based on neaScope scattering type near-field optical microscope (Attocube, Haar, Germany). The beamline radiation is coupled to the near-field spectroscopy module of the microscope, which is based on an asymmetric Michelson interferometer, with an atomic force microscope (AFM) placed in one arm of the interferometer (Knoll & Keilmann, 1999; Hillenbrand *et al.*, 2002; Keilmann & Hillenbrand, 2004; Amarie *et al.*, 2009; Bechtel *et al.*, 2020). The incoming collimated radiation is focused with the parabolic mirror (N.A. 0.46, f=11 mm) on the tip of the AFM probe and the interferogram is detected by a liquid-nitrogen cooled MCT detector that has an active area of 50x50 μm$^2$ and a cut-off at ~ 625 cm$^{-1}$ (Infrared Associates, FL, USA). The path of the beam from the vacuum system to the microscope and the microscope itself are purged with dry N$_2$ and the microscope is protected by an acoustic enclosure. The current configuration of the microscope allows to perform measurements in the spectral range of 600-2000 cm$^{-1}$, determined by the detector, beam splitter, AFM probe and the incoming synchrotron radiation. The IR synchrotron light can be used for both imaging and point spectroscopy experiments. Using the broadband synchrotron light the imaging is performed at the white light position of the interferometer and the observed contrast in the recorded images does not contain exact spectral information related to a specific band, but rather represent intensity changes over the whole broad spectrum. The point spectroscopy method should be used to reveal the spectral changes at the points of interest. In addition to the synchrotron-based modes, the end-station allows to perform pseudo- heterodyne IR-imaging using an additional tunable single-frequency laser source.

The optical amplitude spectrum measured from a Si reference sample and the broadband synchrotron radiation are shown in Figure 3. It is known that the contribution from the near-field signal to the overall signal detected by the scattering type near field microscope increases with the number of the harmonic of the modulation frequency. At least registration of the 2$^{nd}$ harmonic of the optical signal is necessary for a sufficient suppression of the background contribution and extraction of the near-field component from the input signal. The intensity of the optical signal is usually the factor limiting registration of the higher harmonic signals. However, one should also consider the bandwidth of the analog to digital converter (ADC) used in the system, that has a cut off at about 1 MHz. This makes impossible the registration of the fifth harmonic of the signal if the resonant frequency of the AFM probes is too high. In our experiments we could detect up to the fifth harmonic of the optical signal from the Si-reference sample by using AFM-probes with a resonance frequency of < 220 kHz.

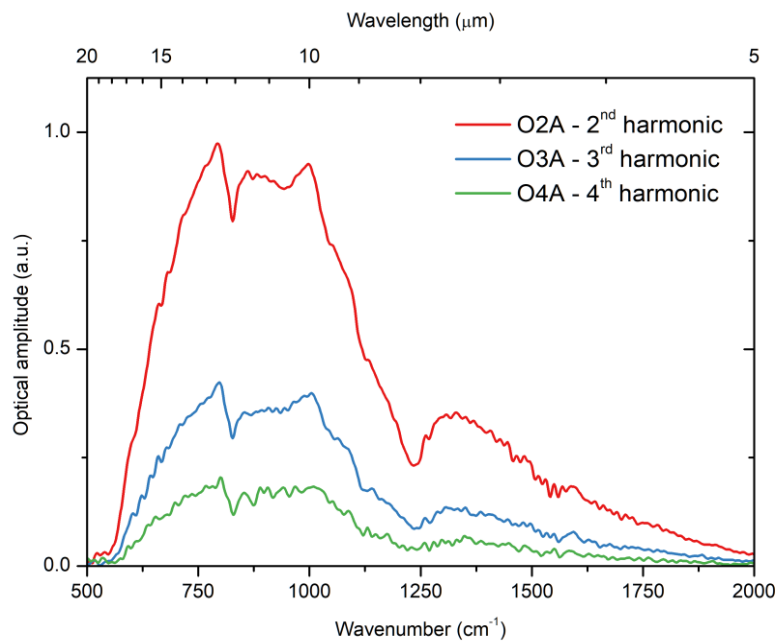

**Figure 3** Different harmonics of the optical amplitude signal collected from the Si-reference sample at the nano-spectroscopy end-station: red/top trace, blue /middle trace and blue/bottom trace correspond to the 2$^{nd}$, 3$^{rd}$ and 4$^{th}$ harmonic of the signal. The spectral resolution of the spectra is about 8 cm$^{-1}$ and the acquisition time was about 15 min. The AFM probe oscillation frequency is 250 kHz and the tapping

amplitude is 75 nm. The high resonant frequency of the AFM probe and the high frequency cut off of the ADC at about 1 MHz does not allow to register the fifth harmonic of the signal.

The spatial resolution of the method is determined by the tip apex radius of the used AFM-probe and can reach a value of < 10 nm (Mastel *et al.*, 2018). The new end-station significantly extends the range of the possible experiments at the beamline, enabling investigations of complex multicomponent materials and hierarchical systems, including energy materials, complex microstructured biomaterials, biological macromolecules such as membrane proteins and protein crystals, hybrid materials, objects of cultural heritage, or minerals with tens of nm spatial resolution. FTIR nanospectorscopy can be used as a single technique or, advantageous in most multi-scale studies, be in combination with the other IR-spectroscopy methods that are available.

In addition to enabling to fuse information from different spatial scales ranging from millimeters over micrometers to nanometers, the acquisition of 'conventional' IR spectra from the same or similar samples also helps the interpretation of the observed near-field spectra – IR spectra collected with the s-SNOM can differ significantly from the much more common FTIR spectra obtained using far-field ATR or transmission methods(Amarie & Keilmann, 2011; Mastel & Govyadinov, 2015).

As an example, recently we investigated orientation of cellulose microfibrils in Sorghum by means of diffraction-limited polarized infrared micro-spectroscopy and a preferential orientation of cellulose and other macromolecules at the different scales ranging from the plant tissues to single cell walls(Veber *et al.*, 2023). The detailed analysis of the anisotropic behaviour of the spectra also allowed to determine the orientation of the cellulose microfibrils indirectly, however averaged over the area of about 5x5 µm$^2$, corresponding to the diffraction limit at 1160 cm$^{-1}$, the wavenumber of a polarization sensitive band in the spectrum of cellulose.

The use of IR nano-spectroscopy technique allows investigations significantly below the diffraction limit and to directly visualize individual cellulose microfibrils as nanoscopic material. Multiple signals are recorded by the system simultaneously. Figure 4 shows images obtained from cotton cellulose microfibrils deposited on a Si-substrate. One can see the topography, the 2$^{nd}$, the 3$^{rd}$ and the 4$^{th}$ harmonics of the optical amplitude signal in Figure 4, panels A, B, C and D, respectively. According to the topography measurements (Z-profile), the width of the smallest microfibril detected in the area is about 30 nm, and its height is less than 2 nm (Figure 4A and 4E, dashed blue trace). The width is comparable to the specified radius of the AFM probe tip apex of 25 nm that used in the experiment. This microfibril can be observed in the optical amplitude signal and is clearly visible in the third harmonic of the IR optical amplitude signal in the white light imaging mode. A larger microfibril of 70 nm width and 12 nm height results in higher contrast in the optical image (Figure 4A and 4E) and can be detected up to the 4$^{th}$ harmonic of the optical amplitude signal (Figure 4D).

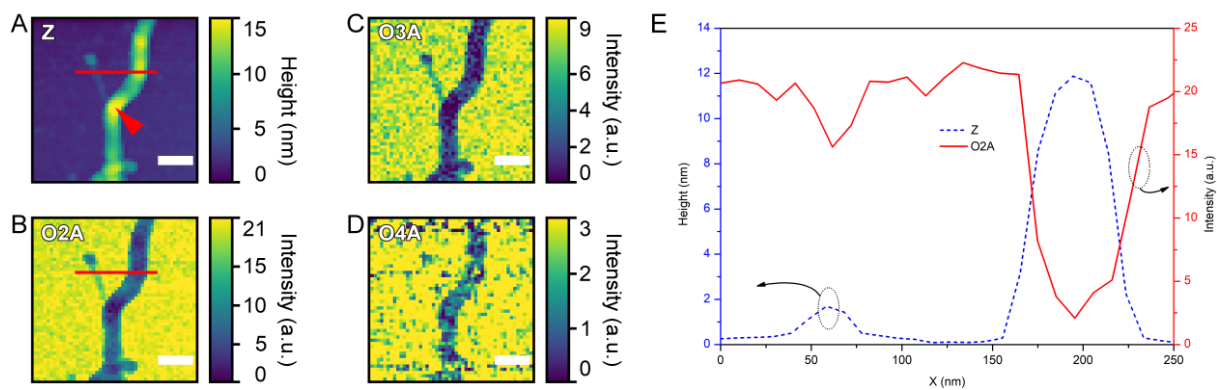

**Figure 4** Z-profile (a) and white light images at second (B), third(C) and forth (D) harmonics of the optical amplitude signal of cotton cellulose deposited on a Silicon substrate. The height and the optical signal profile (E) are taken along the red line shown in panels (A) and (B). The arrowhead in panel A points to the position from where the nano-FTIR spectrum (cf. Figure 5) was collected. The scale bar in panels A-D represents 100 nm. The AFM probe oscillation frequency and the tapping amplitude in the experiment were 259 kHz and 25 nm, respectively.

The typical integration time per pixel for the imaging mode is about 20-50 ms, allowing quick acquisition of images of rather large sample areas. Despite the observed contrast, these images do not contain exact spectral information and represent intensity changes over the whole broad spectrum reaching the sample, different from the typical chemical imaging that uses single spectral bands or a single frequency radiation source. To reveal the specific spectral changes at the point of interest, an interferogram is recorded and converted to the spectrum using the FFT algorithm. The corresponding spectrum of the cotton microfibril from an effective volume of about 30x30x12 nm$^3$ is shown in Figure 5 (bottom trace). Figure 5 also contains near-field (middle trace) and far-field (top trace) absorption spectra acquired from a thick layer of randomly oriented crystalline nano cellulose deposited on a Si-surface. The spectra collected with s-SNOM technique and standard far-field FTIR spectroscopy are in good agreement. Despite slight frequency shifts of the vibrational bands, the good correlation is also observed between the near-field amplitude and the far-field reflection as well as between the near-field phase and the far-field absorption spectra (see supporting information Figure S1). This provides an evidence that many of the vibrational modes of cellulose originate from weak oscillators, and the comparison of nano-FTIR and far-field FTIR absorptions is applicable (Govyadinov *et al.*, 2013), which significantly simplifies the interpretation of the stand-alone near-field spectra in this case.

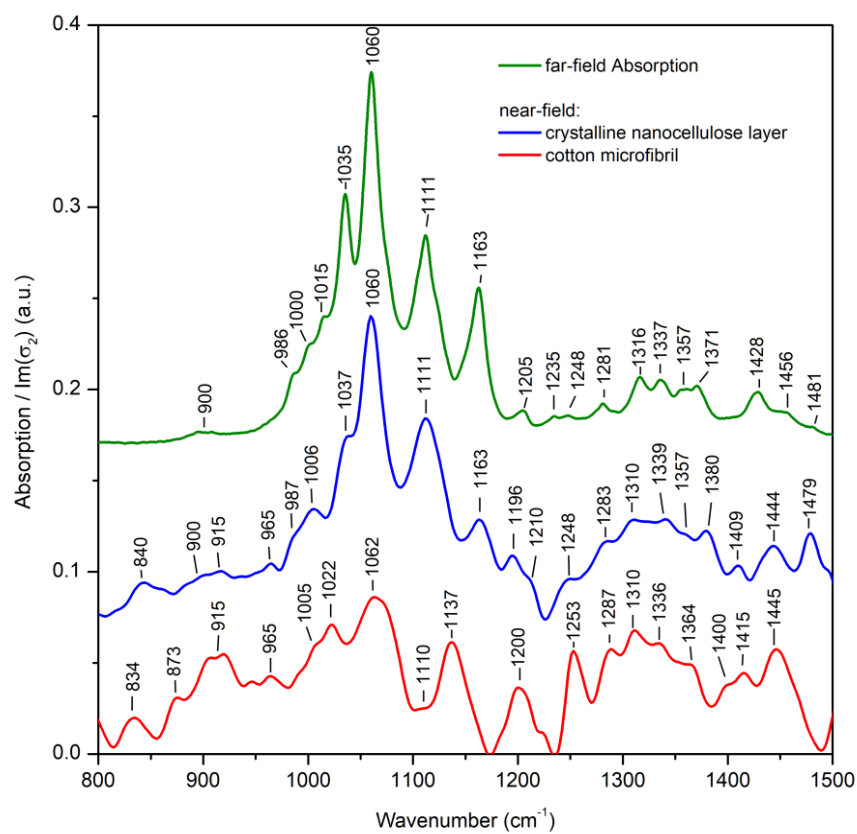

**Figure 5** Near-field absorption spectra of the single cotton microfibril (red/bottom trace) and of the thick layer of unoriented crystalline nanocellulose (blue/middle trace) as well as far field absorption spectrum of the unoriented crystalline nanocellulose layer (green/top trace). AFM probe oscillation frequency and tapping amplitude were: for the cellulose single microfibril 259 kHz and 25 nm; for the crystalline nanocellulose 248 kHz and 60 nm. The vertical scale for the near-field absorption spectra is the same, whereas far-field absorption spectra was scaled arbitrarily to match the intensity of the near-field data. The spectra are shifted vertically for clarity.

At the same time, the vibrational bands present in the cellulose microfibril spectrum (Figure 5, red/bottom trace) differ significantly from the anisotropic, thick reference cellulose sample (Figure 5, blue/middle trace). It is known that the tip-enhanced s-SNOM method is very sensitive to the anisotropy of the sample, due to its higher sensitivity to out-of-plane vibrational modes(Muller *et al.*, 2016). In the experiment here, the single microfibril is oriented normal to the AFM probe, i.e. the relative contribution of the modes normal to and in the direction of the cellulose chain should increase and decrease, respectively, when compared to the unoriented reference crystalline nanocellulose sample. This could indeed explain the absence of the characteristic peak at 1163 cm$^{-1}$ in the single fibril spectrum, since this band is ascribed to the asymmetric C-O-C stretching vibration in the glycosidic linkage and directed in plane in our s-SNOM experiment. Likewise and in accord with this interpretation as well, the bands at 1200, 1253, 1310, and 1336 cm$^{-1}$, ascribed to the C–O–C symmetric stretching of the glycosidic linkage, the CH$_2$ wagging mode, an antisymmetric C-H deformation, and the OH in-plane deformation, respectively (Tsuboi, 1957; Liang & Marchessault, 1959), correspond to the out of plane modes in our experiment, resulting in the higher relative intensity at these frequencies in the single fibril spectrum. Similar change in intensities of the bands listed above is observed in the far-field polarized absorption spectra of oriented cellulose. In supporting information Figure S2 one can see the IR absorption spectra recorded from an oriented cellulose microcrystal upon change of the crystal orientation (supporting information Figure S2). Nevertheless, some difference between the two s-SNOM spectra cannot be explained using this simple concept, e.g. the peak at 1111 cm$^{-1}$ does not demonstrate significant dichroism in macroscopic cellulose sample, but it is absent or low in intensity in the near-field cellulose microfibril spectrum (cf. Figure 5 and Figure S2). It is important to note, that the near-field spectra recorded for different microfibrils and at different points of the same microfibrils demonstrate quite high variability (supporting information Figure S3). This in agreement with other recent work on cellulose microfibrils investigated by means of s-SNOM technique (Kotov *et al.*, 2023). The variability in the observed spectra can be explained by the heterogenous organization of the microfibrils at the ~30 nm spatial scale, in particular local change of cellulose orientation within the microfibril. Further extensive work is currently being conducted to verify this by means of the IR s-SNOM method, however.

**4. Optimization of the beamline performance**

The spectral range of the beam delivered to the end stations is a function of the emission of the light source, the acceptance angle of the M1 mirror, the slot size of the mirror and the transmission function of the beamline optics. The maximum flux after the M1 mirror assuming no slot in the mirror increases monotonously with the frequency of the emitted light (Figure 6). However, the slot allows the high energy X-ray and UV radiation to pass thorough the mirror towards an absorber behind the mirror. Thus the slot is important to decrease the thermal load on the M1 mirror. Since the IR-beamline was commissioned shortly after the start of the synchrotron operation the slot was set to 6 mm, to protect the mirror from possible instabilities of the electron beam positions and the subsequent hit by the high energy radiation in such a case. However, the current configuration significantly decreases the flux at shorter wavelengths, so that, e.g., only about 50% of the radiation from the maximal power is transmitted at λ=2.5 μm. To date we have gained thorough information about the beam stability and the slot size can be decreased down to 2 mm, which will significantly increase the flux at the wavelengths < 10 μm. This modification is currently at the preparatory stage.

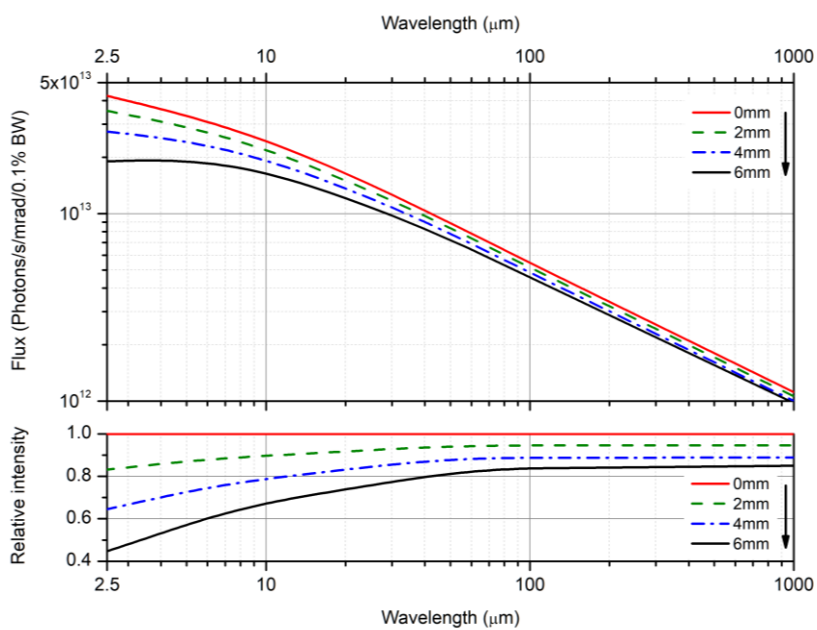

**Figure 6** Flux after the M1 mirror versus wavelength calculated for different slot sizes

The spectral response at the different end-stations can also be modified by utilizing of an appropriate infrared detector. The same detectors can be used at FTIR spectroscopy and microscopy end-stations, in particular Si-bolometers are used to perform measurement if the far-IR spectral region. Unfortunately, the long response time of commonly used Si- as well as transition edge superconducting bolometers does not allow to use them for s-SNOM technique, where oscillations with frequencies up to 1 MHz need to be detected. Currently we are working of optimization and coupling of a liquid-helium cooled Ge:Cu fast photoconductive detector to the nano-spectroscopy end-station, which will ensure sensitivity in the 300-600 cm$^{-1}$ spectral range(Khatib *et al.*, 2018). Moreover, a higher band-gap MCT detector will be added to the system to increase the signal in the range of 1500-4000 cm$^{-1}$.

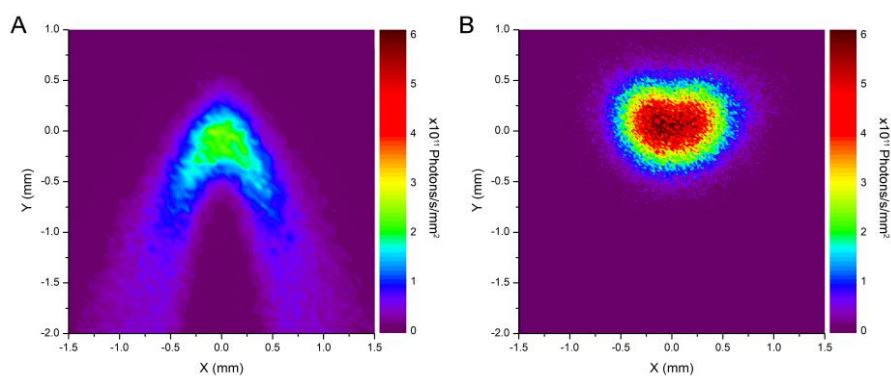

**Figure 7** Calculated beam profile at λ=10 μm for the current beamline optical scheme after M5 mirror (A) and after using potential low-aberration optical scheme, when the beam is scaled to the same dimensions as in the case of (A).

During the upgrade we also considered the possibility to modify the front-end of the beamline to improve the beam quality. The front-end of the beamline uses a traditional source refocusing concept. This optical scheme cannot completely compensate the aberrations caused by the depth and the circular source shape resulting of electrons trajectory (Moreno *et al.*, 2013) and results in a non-circular shape of the beam in focus (see Figure 7A). This can be corrected using a so called magic mirror(López-Delgado & Szwarc, 1976) or a more recent optical outline proposed by Moreno et al. (Moreno *et al.*, 2013).

Our ray tracing calculations show that an optical scheme made according to the Moreno concept applied to the IRIS beamline would result in a much more circular beam profile in focus, and the central spot maximal intensity increases by about 2.5 times (see Figure 7B). The parameters used for the ray tracing are described in the supporting information section "Moreno concept considerations", the 3D model and a side view of the proposed low-aberration optical layout is shown in supporting information Figure S4.

Despite the significant benefit in the beam quality, the realization of the low-aberration port cannot not be done by simple exchange of the mirrors but would require a major modification of the beamline front-end. Thus the adapted Moreno concept could be used as a basis for design of the new IR beamline at the planned 4th generation BESSY III synchrotron (Goslawski *et al.*, 2022), but will not be implemented at the current IR beamline.

The acquisition time of a single interferogram using the nano-spectroscopy end-station depend on many factors, in particular: the sample origin, the AFM-probe tip radius, the probed volume/thickness of the sample, the desired resolution, the spectral range to mention a few. For biological samples, like the single cellulose microfibrils discussed above, acquisition of a good quality spectrum could require up to 0.5-1 hour of integration time. This requests extreme mechanical stability of the AFM microscope positioning system. Our tests show that the drift of the AFM varies in the range of 1-5 nm/min (supporting information Figure S5a), which complicates the accumulation of the spectral data from a single spot for a very long time. Temperature stability of the system is at least one reason for this behaviour. Typical daily temperature variations of our system do not exceed 1 K (supporting information Figure S5b). To overcome the shift issue one can perform series of alternating rather quick interferometer scans with the quick imaging of the sample area of interest. The images can be used to track the shift of the sample and correct the position of the AFM probe for the next interferogram scan. The determination of the shift can be done by using any of the recorded AFM-signals, that results in a high contrast image, by calculating the cross-correlation between two subsequent images. We found that such an algorithm allows to determine the shift with precision better than 1 pixel between the two subsequently recorded images and that mechanical amplitude and phase signals usually suit well for this purpose. Currently we are working on the integration of the software AFM-drift correction into the control software of the near-field microscope.

Recently it has been shown that the performance of the FTIR microscope can be significantly improved by optimization of the beam profile at the end-station using a deformable mirror(Kalkhoran *et al.*, 2022). We believe that this concept can also be applied to improve the coupling of the synchrotron beam to the AFM-probe of the s-SNOM. For a comprehensive vibrational spectroscopic imaging of identical samples in the identical microscope, currently we are developing a combined Raman/IR microscope, which will open further possibilities in multimodal microscopy at the beamline.

## 5. Summary

During the upgrade, the infrared beamline IRIS at the BESSY II storage ring was extended and now it consists of four stationary end-stations: infrared 1) macro-, 2) single shot time-resolved, 3) micro- and 4) nano-spectroscopy. In this summary we show the current outline of the beamline and discuss the experimental methods that are available. Overall, the imaging and spectroscopy capabilities have been improved by enabling a higher imaging rate using an FPA detector for diffraction limited IR microscopy and by extending the spectral range of the diffraction limited experiments to the far-IR spectral region. The nano-spectroscopy end-station is based on a scattering type scanning optical microscope, which enables IR-imaging and spectroscopy with a spatial resolution better than 30 nm. The performance of this new end-station is exemplarily shown here on example of single cellulose microfibrils. All the end-stations are available for national and international user groups. We are constantly developing the beamline to ensure the availability of state-of-the-art IR-spectroscopy techniques to the users.

## 6. Sample preparation and data analysis

The sample for an investigation of individual cellulose microfibrils was prepared using a cellulose nanofibrils slurry (3 wt.%) from cotton (Cellulose Lab Inc., Canada), which was diluted with water to a nominal concentration of $5 \cdot 10^{-4}$ wt.% and drop casted onto a clean Si-substrate and dried in the air.

Purified crystalline Iβ nanocellulose (NAVITAS, Slovenia) was used to prepare reference cellulose samples. A thin layer of this material was obtained by blade casting of the cellulose slurry on a Si substrate resulting in the typical thickness of the film of about 3 μm. This sample was used to record far-field FTIR transmission and near-field SNOM reference spectra of cellulose. Repetitive blade casting was done to get an optically thick layer of the cellulose for a far-field FTIR reflection measurement.

Pt-coated Si AFM tips with an apex radius of ~25 nm (Arrow NCPt, Nanoworld) were used for the near-field IR imaging and spectroscopy.

The Gwyddion (Nečas & Klapetek, 2012) open source software was used for the analysis and the visualization the images obtained by the s-SNOM. The inteferograms recorded by the s-SNOM were processed using an in-house developed script in SciLab open-source software using three-term Blackmann-Harris apodization function window and zero-filling factor of four. Raytracing of the beamline optical path was done with SpotX software (Moreno & Idir, 2001). The plotting of the data was done in OriginPro (OriginLab, Northampton, MA, USA) software.

**Acknowledgements** We thank HZB-BESSY for the allocation of beam time at beamline IRIS. Funding by the German Federal Ministry for Education and Research (BMBF) project 05K19KH1 (SyMS) and by Germany's Excellence Strategy – EXC 2008 – 390540038 – UniSysCat is gratefully acknowledged.

# Supporting information

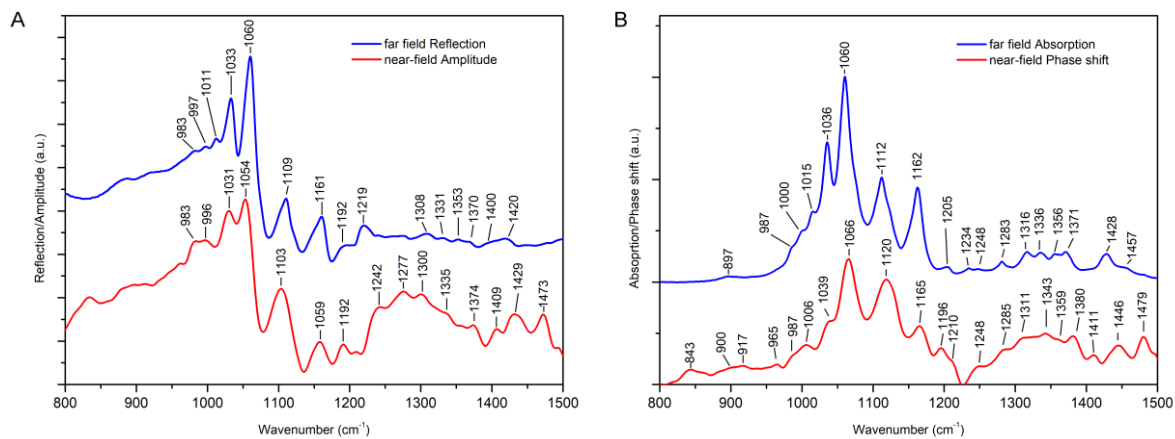

**Figure S1** Comparison of the crystalline nano cellulose far field absorption and reflection spectra with near-field amplitude and phase signal spectra. The second harmonic of the near-field optical signal was used for the amplitude and phase spectra.

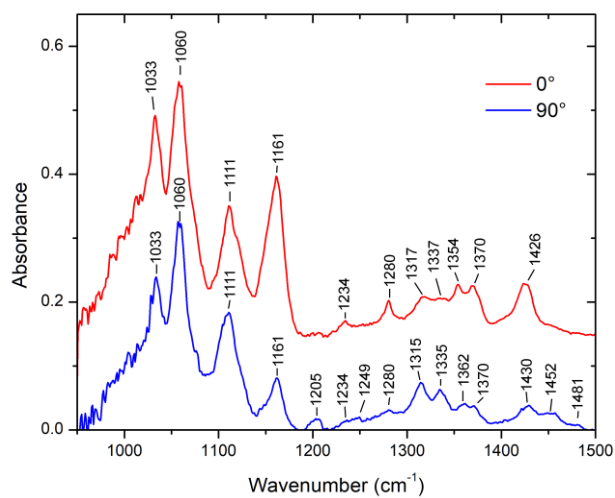

**Figure S2** Polarized far field absorption spectra recorded from a highly oriented cellulose single microcrystal. The infrared light was oriented along and perpendicular to the cellulose chain axis at 0° and 90° spectra, respectively. The zero-degrees spectrum is shifted vertically for clarity. The spectra are reproduced from our recent work (Veber *et al.*, 2023).

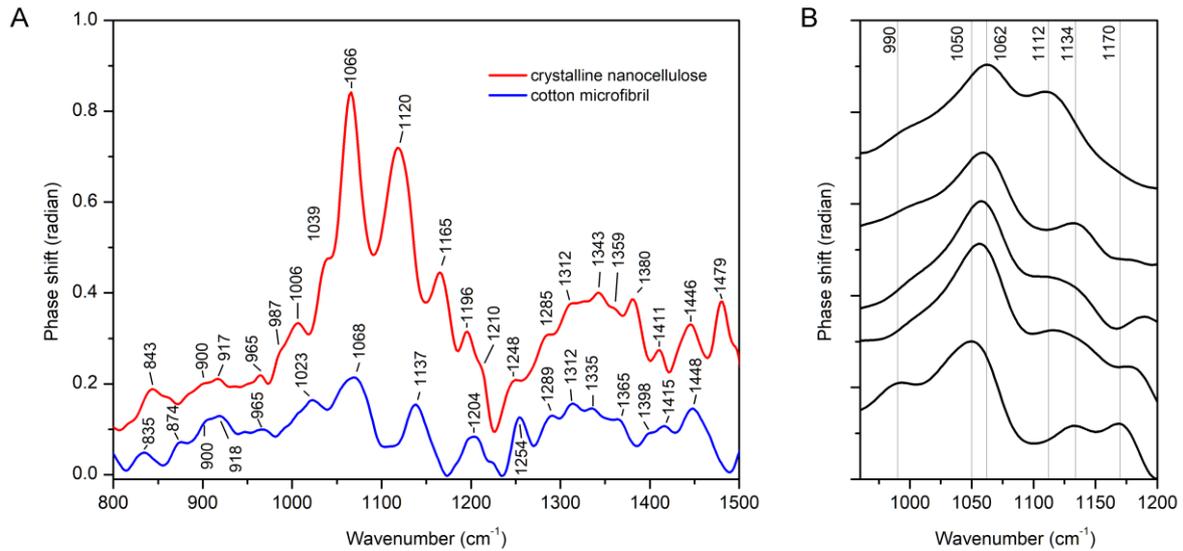

**Figure S3** A) Near-field phase spectra of thick layer of unoriented crystalline nanocellulose (red/upper line) and single cotton microfibril (blue/bottom curve). B) Near field phase spectra collected from different cotton microfibrils. The second harmonic of the near-field optical was used for the spectra.

**S1. Moreno concept considerations**

The Moreno concept allows to solve the depth and the circular source aberration by use of a conical mirror and a long focal length cylindrical mirror. The exact focal length and shape of the mirrors are unambiguously determined by the bending magnet source parameters, position and orientation of the mirrors. To compensate the aberrations all these parameters have to satisfy a set of equations (Moreno, 2017).

For the ray tracing calculations shown in this work the cylindrical mirror M2 was replaced with a conical mirror with the radius of 1910 mm at the center and the radius gradient of 1.9 mm/mm, the position and the orientation of the mirror did not change. The Moreno style scheme cannot be implemented by simple replacement of the current M3 to a new long focal length cylindrical mirror, since the current position of M3 is too close to the dipole magnet. To implement the aberration free outline current cylindrical M3 mirror can be replaced with a flat mirror and an additional cylindrical mirror (M*), determined in terms of the Moreno concept, is installed at the distance of about 3 m from the source, and M* could redirect the beam towards the storage ring wall (instead of the roof), where a new port would be required. At this position and the AOI of 45° M* need to have a focal length of about 15 meters. To decrease the focal length of M* mirror the AOI on both new flat M3 and M* can be decreased to 35°, which results in M* radius of 6670 mm and the intermediate focus at about 5.2 m from M*. The 3D model and side view of the proposed low-aberration optical layout is shown in supporting information Figure S4.

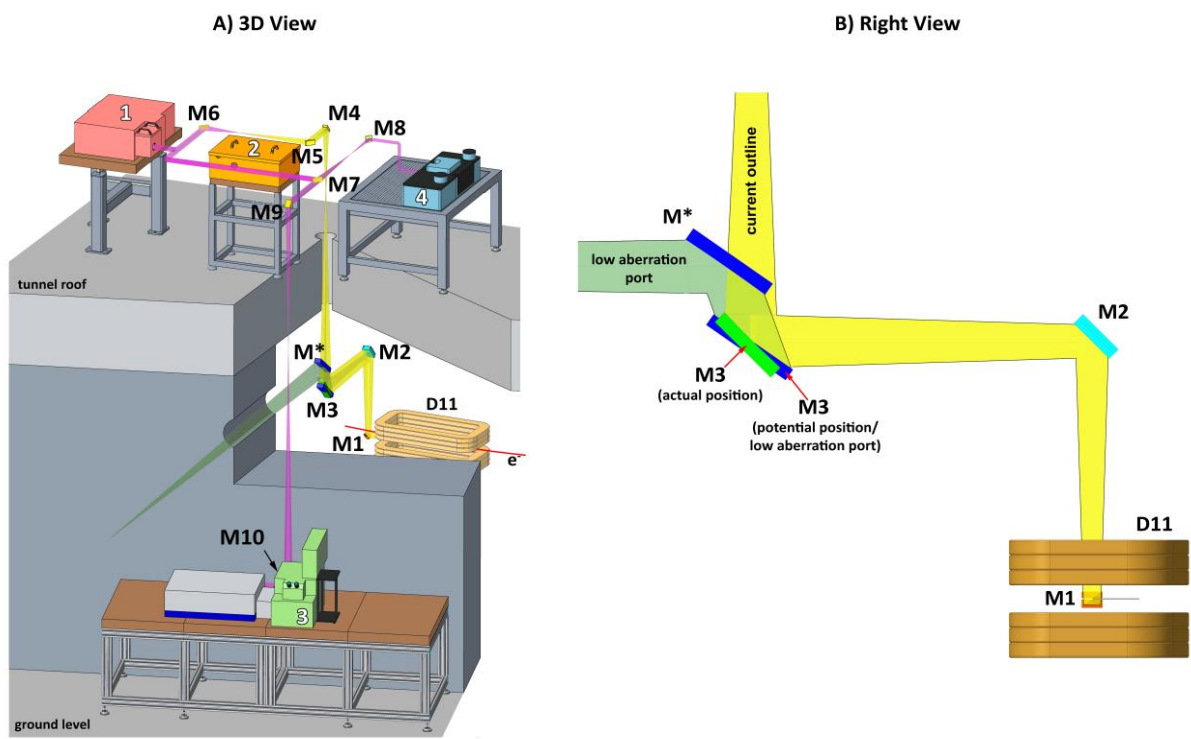

**Figure S4** Outline of potential implementation of the Moreno style low aberration optical port at the IRIS beamline: A) 3D model and B) Side view of the front-end (M1-M3 mirrors) of the IRIS beamline optical scheme. The yellow beam, redirected upwards with M3 (actual position) corresponds to the current optical scheme, the light green beam, redirected (sidewards) with M*, corresponds to potential low aberration extraction port.

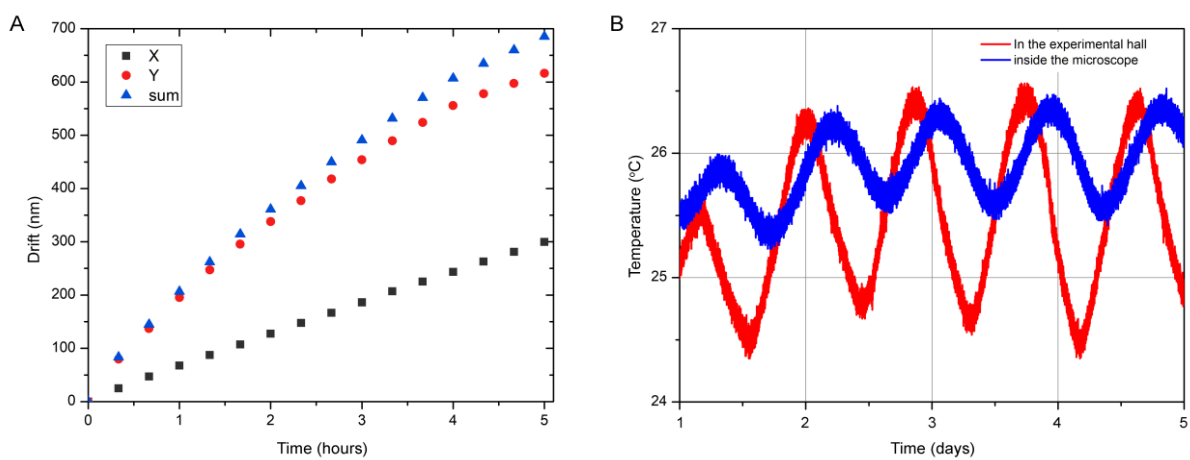

**Figure S5** A) Drift of the AFM probe tip position is time in horizontal (X), vertical (Y) direction and the total value (sum). B) Variation of temperature in the experimental hall of the BESSY II storage ring (red curve) and inside the enclosure of the nano-spectroscopy end-station (blue curve). The day-night temperature variation in Berlin was about 10 °C in the days of the temperature logging.